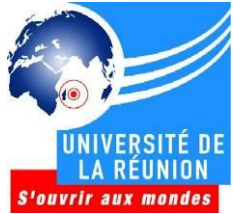
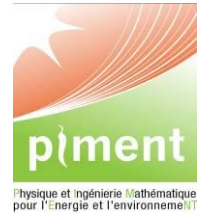

# Submission of manuscript to Energy and Buildings

## Evaluation of the thermal resistance of a roof-mounted multi-reflective radiant barrier for tropical and humid conditions:

*Experimental study from field measurements*

Frédéric MIRANVILLE, Ali Hamada FAKRA, Stéphane GUICHARD, Harry BOYER,

Jean-Philippe PRAENE and Dimitri BIGOT

Contents:

- *Manuscript*

Corresponding author:

**Pr. Frédéric MIRANVILLE**

Physics and Mathematical Engineering Laboratory for Energy and Environment (PIMENT)

Universiy of Reunion

117, rue du Général Ailleret

97430 Le Tampon

tél : 06 92 29 44 87

fax : 02 62 57 95 40

email : frederic.miranville@univ-reunion.fr

# Evaluation of the thermal resistance of a roof-mounted multi-reflective radiant barrier for tropical and humid conditions:

## *Experimental study from field measurements*

Frédéric MIRANVILLE, Ali Hamada FAKRA, Stéphane GUICHARD, Harry BOYER,
Jean-Philippe PRAENE and Dimitri BIGOT

*University of Reunion, Physics and Mathematical Engineering Laboratory for Energy and Environment (PIMENT)*
*117, rue du Général Ailleret 97 430 Le Tampon, France.*
*Phone (+692) 29 44 87, Fax (+262) 57 95 40, email : frederic.miranville@univ-reunion.fr.*

**Abstract**

This paper deals with the experimental evaluation of a roof-mounted multi-reflective radiant barrier (MRRB), installed according to the state of the art, on a dedicated test cell. An existing experimental device was completed with a specific system for the regulation of the airflow rate in the upper air layer included in a typical roof from Reunion Island. Several experimental sequences were conducted to determine the thermal resistance of the roof according to several parameters and following a specific method. The mean method, well known in international standards (ISO 9869 - 1994) for the determination of the thermal resistance using dynamic data, was used. The method was implemented in a building simulation code in order to allow the determination of the thermal indicator automatically. Experimental results are proposed according to different seasonal periods and for different values of the airflow rate in the upper air layer.

**Nomenclature**

| | |
|---|---|
| Tse | Temperature of the exterior surface of the roof |
| Tsi | Temperature of the interior surface of the roof |
| $\varphi$ | Heat flux density through the roof |
| $\varepsilon_1$ | Percentage of difference between the thermal resistance calculated using the entire series of data and the resistance calculated using the database minus one day |
| $\varepsilon_2$ | Percentage of difference between the resistance calculated using the first 2/3 of the series of data and the resistance calculated using the last 2/3 of the database |
| $\dot{Q}$ | Airflow rate |
| $P_c$ | Percentage of reduction of heat flux through the roof |
| R | Thermal resistance of the roof |

# 1. Introduction

## 1.1. Passive design of buildings

Passive design is of great importance for reducing energy consumption of buildings and emission of greenhouse gasses. This important step in the whole process of building construction is more and more improved and relies on a more and more complete set of materials and technical solutions. Among them, insulation products are well known and play an important role in the whole behaviour of the building as well as for the global performance of the building envelope. Mineral wools are well used products for example and are part of mass insulation category. Other types of products exist, as reflective insulation membranes, whose thermal performance is more and more studied [1-3].

The set of products in the field of reflective insulation have greatly evolved during the past fifteen years, and now includes multifoils materials. Although their principle of action is closely linked to the radiative properties of their surfaces, the addition of several layers of foam and wadding as well as reflective foils inside the insulation have an impact on their performances and, consequently, on the global performance of the wall inside which they have been inserted.

Nevertheless, the precise determination of the thermal performances of such technical solutions is difficult, because of their mode of insertion in buildings. Indeed, their principle of action requires the presence of air layers, at each sides of the reflective insulation, in order to induce heat transfer by infrared radiation. Such an assembly, where homogeneous and inhomogeneous layers are mixed can be qualified as complex and complicates the determination of indicators of performance. The configuration of the air layers in particular, opened or closed, naturally or forced ventilated, is fundamental for the effective intensity of heat exchanges across the insulated wall.

Although a lack of results can be noted, it is important to be able to characterise the impact of the insertion of multi-reflective insulation products in buildings. For this, several options can be chosen among which experimental and numerical approaches. For the specific conditions taking place in Reunion Island, characterised by a tropical an humid climate, with strong impacts on buildings components, this problematic is of great importance and requires a recognised methodology, in order to determine both the thermal behaviour and performance of walls equipped with such kind of insulation products. To reach this double objective, both experimental and numerical studies have been conducted. In this paper, the focus is done on the experimental part.

## 1.2. Literature Review

There have been many international studies published about reflective insulation, with the scope of determining their thermal behaviour when inserted in building walls and with the aim of reaching their thermal performances, through a thermal resistance. Moreover, many parametric studies have been proposed, whose purpose was to put in evidence their thermal performances, according to:

- *The location where they are installed*

- *The rate and type of ventilation of the air layers surrounding them*

- *The effect of settling dust*

- *The effect of humidity*

Most of these studies come from the United States and usually consider an attic, either ventilated or not, featuring a nominal thermal insulation using mineral wool, characterised by a thermal resistance R [4-7].

All these studies constitute an important basis for the understanding of the thermal behaviour of reflective insulation and have more recently evolved to point out the thermal resistance of attics under field conditions, using standard methods [8]. The issue of reflective insulation or radiant barrier is still important, especially when dealing with summer comfort conditions under strong climates. The most recent publication indicate that their problematic is more and more studied in Europe, where the number of products distributed for use in buildings is very high and still growing [9]. Theses technical solutions are viewed as an interesting means of insulating buildings, while maintaining the quality of ambient air and the level of insulation in time.

Both numerical and experimental studies are available, with a common point being exposed in almost all publications: the scale of modelling of radiant barrier in building is most likely the multizone one, with a nodal description. Moreover, two categories of tools are developed to predict their impact on buildings, specific ones and generic ones. In the last case, the modelling of radiant barriers is generally qualified as integrated, in the sense that it is described using the existing concepts that were used when developing the code [10]. Specific models are on the contrary based on given configurations [11].

### 1.3. The French context

As indicated in previous studies, reflective insulation is considered in France as a separate means of insulation. Unlike the English-speaking countries, where radiant barriers are used in combination with mass insulation, reflective insulation technology is opposed to mass insulation and is subject to a virulent debate about their performances. Two sets of opponents are facing each other, those of the distributors, showing that reflective insulation is equivalent to 10 to 20cm of mineral wool, and the regulators ones, much lower. Moreover, classical experimental methods of determining the thermal resistance of reflective insulation are not adapted to the correct evaluation of their performances, in particular due to the steady state conditions. In actual conditions indeed, the dynamic behaviour of reflective insulation technology is of great importance on the impact of the heat flux rate through the considered building component [12].

Besides, the technology of reflective insulation in France and more generally in Europe have evolved, and products distributed under the name of multi-reflective radiant barriers (MRRB), are now proposed (see Figure 1). These products are characterised by a large number of intermediate layers between the low emissivities faces, most often constituted with foam, wadded and even containing additional reflective foils. In the extension of the previous problematic, the thermal performances of such evolved products have to be determined, both from experiments and numerical studies.

*Figure 1: Multireflective insulation as the new technology for reflective insulation*

## 2. Problematic and methodology

### 2.1. Introduction

Reunion Island is part of the French overseas departments, located in the Indian Ocean, and characterised by a tropical and humid climate. In summer in particular, temperatures are usually well above 30°C, and the rate of relative humidity is generally above 80%. Moreover, wind conditions are generally weak, with wind speed oscillating between 0 and 1.5 $m.s^{-1}$. Finally, it is important to outline that this location is often subject to cyclones, which can be very powerful and generate important disasters.

Architectural design in Reunion Island has evolved rapidly from a very traditional construction method to an imported mode of construction of building, mainly from France. The consequence of this rapid evolution is the presence of several types of residential buildings, some of them constructed with wood, others being concrete enclosures and more recent ones combining concrete and traditional art of construction. For this last category of houses, more and more spread in the territory, the roof is mainly composed of corrugated iron, with a framework being composed of galvanised stainless steel. The ceiling is usually composed of plasterboard, without any installed insulation product. These types of roofs are subject to important heat gains, and due to the high temperature reached by the roof coverings, of the order of 80°C in summer, the radiative load from this part of the roof is non negligible. For such configurations, the use of reflective insulation is indicated, as published in [2].

Very recently, a thermal regulation has been developed and put in application for Reunion Island [15]. This constitutes a major step in building design in this region and is intended to promote the use of insulation and other technical solutions to minimise the energetic consumption of buildings. In such a context, it is important to reach thermal indicators of the performances of multireflective insulation for typical roofs of Reunion Island.

### 2.2. Problematic

The insertion of a MRRB in a typical roof from Reunion Island induces complex physical phenomena. In particular, the thermal modes of heat transfer are fully combined and coupled. Due to the presence of air layers, heat is transferred from the roof covering to the radiant barrier through convection and radiation, to the first side of the product to the other side both by radiation and conduction, and from the lower side to the ceiling also by convection and radiation. Contrary to the insertion of a mass insulation product, which mainly generates heat transfer by conduction and convection, radiation has to be taken into account in the case of reflective insulation, both outside and inside the product. This constitutes a major difficulty when dealing with such kind of technical solutions, and has to be treated very precisely.

Secondly, important parameters are of importance on the whole performance of the thermal system. The radiative properties of the low emissivity surfaces of course, but also the ventilation rate of the air layers included surrounding the MRRB. Usually, although two air layers are installed when a radiant barrier is inserted into a roof, only one is ventilated, the upper one. In actual cases, the upper air layer is naturally ventilated, but in some situations, it can be forced-ventilated. The conditions of ventilation of the upper air layer are thus important parameters to monitor when dealing with thermal performances of multireflective insulation products.

Thus, our problematic can be summarised as follows: determining the thermal performances of MRRB, installed in typical roofs from Reunion Island, according to important parameters like seasonal effects and, in particular, the rate of ventilation of the upper air layer of the assembly.

## 2.3. Methodology

### 2.3.1. Overview

Answers to the previous problematic are rather complex and our choice for trying to put in evidence validated elements is based on a combined approach with both numerical and experimental steps.

For this, a dedicated building simulation code was developed and a specific experimental platform was set up. Publications have already been done on these tools, designed and developed in Reunion, and the interested reader is invited to see [2], [10] and [13] for details.

The focus of this paper is the determination of the thermal performance of multireflective insulation, in field conditions and for a realistic configuration. For this, expected results are the thermal resistances of the wall in which the radiant barrier has been inserted, from dynamic field measurements. To be able to put in evidence such results, it was important to select an appropriate method, and to include it in our tool. The ISO 1994-9869 [14] standard was thus chosen, and more precisely the mean method, to be able to determine the equivalent thermal resistance of the roof. To test several configurations, the previous method was even extended, as explained in the following paragraph.

### 2.3.2. The mean method

The mean method is well known for the calculation of thermal resistances of building elements from dynamic measurements and is thus an interesting approach to assess results from field values. The method is fully described in the international standard ISO 9869-1994, and is based on the following simple equation, applied to series of dynamic data:

$$R = \frac{\sum_{i=1}^{n}\left(T_{se,i} - T_{si,i}\right)}{\sum_{i=1}^{n}\varphi_i}$$

Where $\begin{cases} R \text{ is the thermal resistance of the wall} \\ T_{si,i} \text{ is the interior surface temperature} \\ T_{se,i} \text{ is the exterior surface temperature} \\ \varphi \text{ is the heat flux through the wall} \end{cases}$

Several conditions are necessary to validate the results, and ensure that the energy balance over an entire period is respected. Considering a series of temperatures (interior and exterior faces) and heat flux (through the wall), the result of the calculation is judged valid in the following conditions:

1. *The percentage of difference ($\varepsilon_1$) between the resistance calculated using the entire series of data and the resistance calculated using the database minus one day is less than 5%*

2. *The percentage of difference ($\varepsilon_2$) between the resistance calculated using the first 2/3 of the series of data and the resistance calculated using the last 2/3 of the database is less than 5%*

When these conditions are satisfied, the resulting thermal resistance has converged toward the value obtained in steady-state conditions.

The method has been implemented in the building simulation code ISOLAB, as a specific module for the determination of the thermal performance of building components. Hence, the user is able to use the module with series of results of simulations as well as series of measurements. It is therefore possible to lead purely theoretical studies, with the resulting evaluation of the thermal performance of the considered building component, or dealing with experimental work. In this last case, the module of the building code is used as a standalone tool.

During the implementation of the method, some parts of the process, anterior to the calculation of the thermal resistance, have been automated. The verification of initial and final bounds of the database is the first point; it is important indeed that the thermal state of the system is equivalent both at the beginning and at the end of the series of measurements. This verification ensure that the principle of conservation of energy is respected on the entire period considered. From a very strict point of view, the thermal resistance is defined for steady-state conditions only, when the thermal system is in equilibrium. To use the mean method to obtain representative results, absolute care must be taken when dealing with dynamic measurements, to ensure that energy is not stored in the system over the considered period of measurements. This point is a key one, which strongly complicates the use of the method when dealing with complex walls. As stated previously, these specific types of building components are constituted by both an assembly of homogeneous and fluid layers, and are subject to combined modes of heat transfer. During a sequence of time, heat is thus stored in the wall and during another one, heat can be released, depending on the sense of the heat flux.

Our approach relative to complex wall is to apply the mean method even when fluid layers are included in the assembly. This is often the time in buildings, and their behaviour can have a great impact on the performance, depending on the convection intensity. When dealing with air layers for instance, which can be ventilated or not, the proportion of modes of heat transfer is greatly modified, and the associated impact is important. To include this important parameter when using the mean method, a specific routine has to be implemented to ensure that the conservation of energy is granted over the studied period. This can be done, from a technical point of view by scanning the input database (coming from field measurements in our case) in order to identify similar ending conditions as the starting ones.

Once this important condition is verified, it is necessary, to validate the final result, to take into account a database which leads to correct values of $\varepsilon_1$ and $\varepsilon_2$. These two indicators are included for verification in the mean method, but, when dealing with time series of data, especially coming from field measurements, they are not necessary verified. To avoid these cases, and to allow a large exploitation of the given database, implementation has been conducted according to the synoptic indicated on Figure 2. This procedure allows to consider different databases, extracted from the initial one, and to obtain validated values.

*Figure 2: Synoptic of the process of calculation of the thermal resistance implemented in ISOLAB*

To take care of the reference database and to increase its potentialities of treatment, some additional capabilities have moreover been integrated, consisting in several filters. Hence, the whole database can be taken into account for the calculation as well as part of it according to the following parameters:

- *Positive heat flux condition*
- *Day-time selection*
- *Night-time selection*
- *User-time selection*

These criteria allow a pre-treatment of the database, before the actual calculation. It increases the possibilities of exploitation of the series of data, often laborious to obtain, to constitute alternative options when the validity of the final results is not obtained.

# 3. Experimental environment

## 3.1. Experimental devices and instrumentation

### 3.1.1. *The experimental platform for building physics research*

The dedicated experimental tool used for this study is part of an experimental platform, installed at the University of Technology of Saint-Pierre, at a low altitude from the sea level (68 m). This area is quite important, more than 600 m² being dedicated to the observation of physical variables relative to building physics. It is composed of different test cells, some of them being low scale devices (named ISOTEST) and another one being a normal scale building (named LGI). A meteorological station is also installed on site, to measure precisely the climatic conditions near the experimental devices. Each of the test cells faces the geographical north, in order to receive symmetrical solar solicitations during the day. No shading occurs from one test cell to another, to ensure that thermal interactions between the cells are negligible.

### 3.1.2. *Details of the LGI test cell*

The LGI test cell is representative of a typical room of a building. It has an interior volume of about $29.8m^3$ and is designed with a modular structure, which allows testing several configurations and phenomena. The walls are movable for this reason. It features opaque, vertical walls, with blind-type windows, a glass door and a roofing complex including an MRRB. The details of the arrangement of these elements are given in Table 1.

*Table 1: Details of the construction of the LGI cell*

It is equipped with a standard roof, installed according to the manufacturers' specifications. In addition, it includes a glass door (with upper and lower panes) and an aluminium blind, as shown in Figure 3.

*Figure 3: The LGI test cell*

The test cell is orientated facing north. The corrugated covering is a dark colour, for the development of an extreme input from the roof. The cell is also equipped with mechanical ventilation and split-system air conditioning. For the experimental sequence, the window panes in the door were masked, as were those of the blind.

For the study of MRRB, a new specific roof was designed and installed. In the section shown in Figure 4, one can see its geometric details. It is composed of a corrugated covering made of galvanised steel (both sides are varnished dark blue), an MRRB and a ceiling made of plasterboard. The framework includes rafters with a C-shaped profile and galvanised steel spacers; it also includes wood rafters, whose assembly form the upper air layer. The lateral faces are made of dark-coloured galvanised steel sheets. The roof is inclined at 20° to the horizontal, which is the angle most frequently encountered in Reunion Island.

*Figure 4: Section and front view of the specific roof installed for the study of MRRB (Multi-Reflective Radiant Barriers)*

Moreover, as the evaluation of the ventilation rate of the upper air layer on the thermal resistance was part of the problematic, a specific device was designed. It is composed of two ventilation boxes, the first one attached to the air input of the upper air layer, and the second one being installed at the end of the air layer. A mechanical ventilation fan is mounted at the base of the cell so that the ventilation rate in the upper air layer can be varied. A schematic of the device is shown on Figure 5, and illustrates in particular the airflow path when the system is functioning.

*Figure 5: Section view and photo of the device for the mechanical ventilation of the upper air layer of the roof of the LGI cell*

### 3.1.3. Instrumentation of the LGI cell

The LGI cell is equipped with approximately fifty sensors, including those relative to the observation of its passive behaviour. These sensors are located both in the enclosure and the roof. The enclosure is equipped with thermal sensors on each side of each wall (north, south, east and west) and the interior air volume has sensors at three different levels from the floor to the ceiling, to put in evidence the effect of the air stratification. Moreover, thermocouples are sealed in the concrete floor of the cell, to allow the determination of the boundary conditions from the ground. Each thermocouple has been verified and is whether disposed on walls for surface temperature measurements, inserted in an aluminium cylinder for air temperature measurements or put inside a black globe for radiant temperature measurements.

The complex roof is also fully instrumented, with surfaces temperatures for the roof covering, the MRRB and the ceiling. Air temperatures are also measured in the lower and upper air layers as well as radiant temperatures using black globes. Heat fluxmeters are installed on each part of the roof, and give access to heat flux transmitted through the roof covering, the MRRB and the ceiling. Specifically for the parametric study of the thermal performance according to the ventilation rate of the upper air layers, hot wire anemometers have been inserted into the upper air layer, for the determination of the airflow speed, and consequently the airflow rate.

Each thermocouple has been calibrated on site and the other sensors were verified in the factories. The absolute error from the thermocouples is estimated to be ± 0.5 °C, and the precision of the heat fluxmeter is 5%. The hot wire anemometers have a precision of about 0.5 m.s$^{-1}$.

All the sensors are connected to a datalogger, installed in the test cell, and the collection of data is automatically done every 15 min; data are periodically saved on a dedicated computer.

### 3.2. Climatic data and experimental sequences

Climatic data for the experimental sequences where measured on site with a dedicated meteorological station. The physical variables observed were solar radiation (global, direct and diffuse, on a horizontal plane), wind speed and direction and temperature and relative humidity of exterior air. Each climatic value is measured every minutes and an average is done each 15 min.

The experimental sequences for this study were both in winter and summer seasons, and data were measured over more than one year. During this period, several configurations of the test cells have been monitored, as indicated in the Table 2.

*Table 2: Configurations of the test cells during the experimental period*

# 4. Results

## 4.1. Introduction

Once the whole experimental set-up is installed, sequences of measurements have been run, for more than one year. LGI test cell were monitored continuously over the experimental period, to ensure quality of measurements and avoid risks of malfunction. Several periods of ten days in average, corresponding to the several configurations indicated previously, have been obtained. These databases were used to determine the R-values, according to the mean method, with the procedure exposed previously.

A summary of the process of exploitation of the measurements is proposed on Figure 6.

*Figure 6: Overview of the exploitation method of measurements*

From each sequence, a calculation of the thermal resistance of the roof was run using the module implemented in ISOLAB.

Moreover, to allow a better understanding of the thermal behavior of the whole test cell, but especially the roof, several curves were drawn for each experimental sequence:

    *1. an overview of the climatic sequences*

    *2. heat flux profile through the roof*

    *3. temperatures difference through the roof (between the roof covering and the ceiling)*

The following paragraphs contain the graphical evolutions of the previous variables for the corresponding seasons. The calculation of the thermal resistance of the roof is done at the end of each paragraph, with the indication of the validity of the results.

### 4.2. Experimental results obtained with the LGI cell

#### 4.2.1. *Sequences in winter*

- Climatic conditions:

Winter season in Reunion is characterised by trade winds, whose impact on the exterior air temperature and on relative humidity can be important. Moreover, depending on the location, climatic conditions can vary in large proportions. At low altitudes indeed, on the coasts for example, difference between summer and winter, depending on wind conditions, can be low, with exterior air temperatures and solar radiation allowing to obtain comfortable conditions. At high altitudes, the situation is very different, and the combination of low exterior air temperatures and high relative humidity accentuate the feeling of cold.

In our case, the experimental platform is located at a low altitude, and experiences mild weather. On Figure 7 are presented the main characteristics of the experimental sequence considered, with exterior air temperature, relative humidity, wind speed and solar radiation. From the evolution of wind speed in particular, it can be seen that, over the period of seven days, only one was subject to trade winds. During this particular day, wind speed reached up to 6 m/s, and accordingly, exterior air temperature was slightly inferior of about 1.5°C. Sometimes, trade winds can last longer and consequently, exterior air temperature is relatively low.

*Figure 7: Climatic conditions for the winter experimental sequence (Direct, Diffuse and Global indicate solar radiation)*

- Natural ventilation

In the case when the upper air layer of the roof of the LGI test cell is naturally ventilated, surface temperatures of the boundaries of the roof (roof covering and ceiling) and heat flux through the roof are proposed on Figure 8. A period of seven days has been chosen for a better observation of the evolutions of the physicals parameters.

From the curve of the heat flux through the roof, it can be seen that the evolution follows a daily cycle, with an average value of 0.4 $W.m^{-2}$, a maximum value of 6.2 $W.m^{-2}$ and a minimum value of -1.2

W.m$^{-2}$. The temperature difference between the roof covering and the ceiling also follows a periodical evolution, with an average value of 3.14°C, a maximum value of 33.58°C and a minimum value of -7.14°C.

*Figure 8: Heat flux through the roof and temperature difference between the two boundaries (Tse: temperature of the exterior surface of the roof – Tsi: temperature of the interior surface of the roof) for the winter period - Natural ventilation case*

From these parameters, it is possible to determine the R-value, from the mean method. Application of the module implemented in ISOLAB code lead to the following results:

$$\begin{cases} R = 6.24 \ m^2.K.W^{-1} \\ \varepsilon_1 = 2.21\% \\ \varepsilon_2 = 4.78\% \end{cases}$$

The result is representative of very good performances of the roof. Nevertheless, when the upper air layer is ventilated, conditions of air motion inside the air layer greatly influence the energetic behavior. This results in some difficulties, sometimes, to apply the mean method, because of the intensity of the convection. When wind speed in the air layer reaches high values, stationary conditions are more and more difficult to obtain and consequently the R-value calculation can fail.

- No ventilation

In the case considered in this part, the upper air layer is obturated and hence, no ventilation takes place. The corresponding curves are indicated on Figure 9, where it is possible to see that heat flux through the roof is slightly higher than in the previous case. The average value is 2.02 $W.m^{-2}$, the maximum 11.44 $W.m^{-2}$ and the minimum 2.36 $W.m^{-2}$. The difference of temperature between the two boundaries of the roof are quite of the same order, with an average of 3.39°C, a maximum of 33.47°C and a minimum of -8.85°C.

Figure 9: Heat flux through the roof and temperature difference between the two boundaries (Tse: temperature of the exterior face of the roof – Tsi: temperature of the interior face of the roof) of the roof for the winter period - No ventilation case

The application of the calculation module leads to the following values:

$$\begin{cases} R = 1.66 \ m^2.K.W^{-1} \\ \varepsilon_1 = 1.20\% \\ \varepsilon_2 = 4.98\% \end{cases}$$

This time, the calculated R-value is much lower than in the previous case, and is more representative of values obtained in steady-state conditions. This can be linked to the intensity of the convection heat transfer in the upper air layer, which, in this case is lower than in previous case. More precisely, the phenomenon was advection in the previous case, whereas in this configuration, the air layer is subject to convection, whose intensity is much lower.

### 4.2.2. Sequences in summer

- Climatic conditions

For the summer period, the climatic conditions are illustrated on Figure 10. Compared to the winter period, wind speed is lower (1.25 m/s in average), and exterior air temperature is higher (25°C in average). Moreover, solar radiation is stronger, and is sometimes more than 1000 $W.m^{-2}$, which is often the case in summer in Reunion. The combination of low air speed and strong solar solicitation often generates situations of overheating in dwellings, in particular ancient ones. This is the main reason for the economic development of active means of cooling like split systems for example. Relative humidity is also very high, what accentuate the feeling of heat.

*Figure 10: Climatic conditions for the summer experimental sequence (Direct, Diffuse and Global indicate solar radiation)*

- Natural ventilation

In this configuration, the curves of heat flux through the roof and temperature difference between the surfaces of the roof are presented on Figure 11. The daily cycle is again observable and curves are comparable to the case without ventilation, in winter, in terms of intensity.

*Figure 11: Heat flux through the roof and temperature difference between the two boundaries (Tse: temperature of the exterior face of the roof – Tsi: temperature of the interior face of the roof) for the summer period - Natural ventilation case*

The calculation of the R-value can be run, using the proposed method, which leads to the following result:

$$\begin{cases} R = 1.44 \ m^2.K.W^{-1} \\ \varepsilon_1 = 1.83\% \\ \varepsilon_2 = 0.14\% \end{cases}$$

This value is comparable to the case without ventilation, in winter. This can be explained because of the conditions of air motion in the air layer, which, according to the lower wind speed, are similar to those of the case when there is no ventilation. In summer indeed, trade winds never happen, and consequently, a regime of breeze of low intensity takes place.

- No ventilation

When no ventilation is imposed to the upper air layer, the resulting curves are given on Figure 12. The regime of heat flux is rather similar to the previous one, as well as the regime of temperatures at each side of the roof.

*Figure 12: Heat flux through the roof and temperature difference between the two boundaries (Tse: temperature of the exterior face of the roof – Tsi: temperature of the interior face of the roof) for the summer period - No ventilation case*

The result of the calculation of the R-value is the following:

$$\begin{cases} R = 1.47 \ m^2.K.W^{-1} \\ \varepsilon_1 = 0.67\% \\ \varepsilon_2 = 4.78\% \end{cases}$$

This result is very similar to the previous one, which confirms the reasons indicated above. Aeraulic conditions of the upper air layer have a great impact on the whole energetic behavior, and advection phenomenon, in the case of a naturally ventilated air layer, leads to high performances. Even if care should be taken for the calculation of the R-value in this last case, the resulting value allows to illustrate the observed good performance.

- Controlled ventilation

For this part, the specific device for the mechanical ventilation of the upper air layer of the LGI cell was used. The system is equipped with an air inlet in the back of the test cell (facing the dominant wind) and an air outlet at the front. Such a system is able to achieve constant values of airflow rates. In the ventilation duct (for extracting air) an anemometer has been set up, thus giving access to the airflow speed in the outlet and also to the airflow rate [16].

To put in evidence the influence of the ventilation on the energetic performance of the roofing complex, several sequences of measurements were carried out, whose objective was to determine the airflow rate in the air layer, from the measurement of the air speed. The result of this procedure is proposed in Table 3.

*Table 3: Results of the determination procedure of the airflow rate in the upper air layer of the LGI cell*

Only three sequences of measurements were made, because of the low resolution of the mechanical ventilation system. To modify the airflow rate indeed, the position of an iris has to be changed. During the tests, only three positions showed a significant difference in terms of air speed in the duct.

Results of flux through the roof structure are presented for each case of forced ventilation. On Figure 13 are shown evolutions of heat flux and surface temperatures through the roof. It can be seen that increasing the airflow rate generates a decrease of the heat flux through the roof, and also a decrease of the temperature difference from one boundary to the other one of the roof.

The percentage of reduction in heat transfer is estimated by the percentage of the ceiling heat flux reduction Pc and can be expressed as:

$$Pc = \frac{\int_{test\ period} \varphi_{without\ ventilation}\ dt - \int_{testperiod} \varphi_{with\ ventilation}\ dt}{\int_{test\ period} \varphi_{without\ ventilation}\ dt} \times 100$$

Using this relation, the percentages of reduction of heat flux through the roof indicated in Table 4 were obtained.

*Table 4: Percentage of reduction of heat flux through the roof, according to the airflow rate of the upper air layer*

*Figure 13: Heat flux and temperatures differences evolutions through the roof, according to the airflow rate in the upper air layer*

Even though the curves of heat flux through the roofing complex between the case without ventilation and the case with natural ventilation, exposed previously, are quite similar the benefit of a ventilated air layer can be put in evidence from Figure 13: the heat flux is reduced.

When an airflow rate is imposed to the upper air layer, the percentage of reduction of the heat flux through the ceiling is very important compared to the case without ventilation. This is due to the aeraulic phenomena which becomes predominant in comparison with thermal phenomena. Consequently a little part of the heat flux through the corrugated iron roof top crosses the ceiling due to the discharge of energetic gain by the ventilated air layer (advection phenomenon).

For each sequence, the thermal resistances of the roofing complex were calculated. The results are indicated in Table 5.

*Table 5: Thermal resistance values according to several airflow rates*

When inserted on a graphic, these values show a linear evolution, as indicated on Figure 14. A linear regression can then be done, to determine the associated mathematical relation:

$$\begin{cases} R = 0.0069\dot{Q} + 1.4528 \\ R : \text{ thermal resistance in } m^2.K.W^{-1} \\ \dot{Q} : \text{ airflow rate in } m^3.h^{-1} \end{cases}$$

A relation between the thermal resistance and the airflow rate in the upper air layer can thus be elaborated. As seen on Figure 14, this confirms the great dependence of the thermal performance of the roof from aeraulic conditions in the air layer, and put in evidence the advantage of MRRB, compared to mass insulation, which usually, completely fill air layers. MRRB not only significantly decrease thermal radiation through the roof but also promotes the use of advection to block the thermal loads in summer.

*Figure 14: Evolution of thermal resistance according to the airflow rate in the upper air layer of the roof*

## 5. Conclusion

Passive cooling of buildings is very important under hot climates, and relies on several technical solutions. Among them, thermal insulation products are more and more used and as a consequence, the need for thermal performance indicators is growing. For mass insulation products, thermal performance is well characterized by the thermal resistance or R-value, well known in regulations for example. On the contrary, reflective insulation products suffer from a lack of robust data, and as a result, many distributors promote their product indicating their own values of thermal performances.

Radiant barrier have recently evolved and new technologies have appeared. Multi-reflective radiant barriers are one of them and are constituted as an assembly of multiple layers of wadding, foam and reflective foils. Their insertion in building components, usually the roof, is classically done between air layers, allowing the transfer to be in majority by infrared radiation. Distributors often promote the ventilation of the upper air layer of the roof, for better performances.

In Reunion, where MRRB are more and more used, a dedicated platform has been set-up to study the thermal performance of radiant barrier, and thus MRRB. With many years of works on the subject, this study has been proposed, in order to assess thermal performance indicators, from field measurements and in realistic conditions. LGI cell have thus been equipped, and more than one year of measurements have been conducted.

From these measurements, the thermal performance of MRRB has been determined, using a well known method, the mean method. It has simply been extended to be used with building components including air layers, by an iterative procedure, ensuring automatically that the principle of conservation of energy is respected. For this, the main contribution was to implement the modification of the database, which is simply decreased progressively during the process, until reaching the validated result.

The method has been tested on LGI cell, both in winter and summer conditions. The results are proposed in Table 6.

*Table 6: R-values from field measurements using LGI cell*

From these values, it can be noted the good correlation between values obtained when upper air layer if not ventilated, both in summer and in winter, and the value obtained in summer, when naturally ventilated. The results indicate that the method is able to determine the thermal performances of MRRB, according to climatic conditions. Moreover, the difference between the results in winter is linked to the aeraulic conditions in the air layer. Due to trade winds, air speed in the air layer is more important in winter, and generates a higher value of the thermal resistance.

To validate the impact of aeraulic conditions in the upper air layer, a sequence with controlled mechanical ventilation of the upper air layer was conducted. It has been shown that a linear relation can be derived from field measurements, translating the link between the thermal resistance and the airflow rate in the air layer. This simple relation could be used by architects and engineers, to evaluate the global consumption of buildings, during the design stage.

This experimental study has been coupled with a numerical part, dedicated to the prediction of the thermal behaviour of buildings including roof-mounted MRRB. The results will be presented in a next publication.

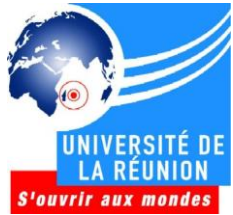

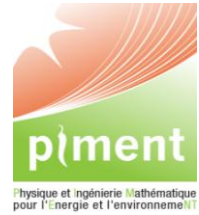

# Submission of manuscript to Energy and Buildings

## Evaluation of the thermal resistance of a roof-mounted multi-reflective radiant barrier for tropical and humid conditions:

### *Experimental study from field measurements*

Frédéric MIRANVILLE, Ali Hamada FAKRA, Stéphane GUICHARD, Harry BOYER,

Jean-Philippe PRAENE and Dimitri BIGOT

Contents:

Corresponding author:

**Pr. Frédéric MIRANVILLE**

Physics and Mathematical Engineering Laboratory for Energy and Environment (PIMENT)

Universiy of Reunion

117, rue du Général Ailleret

97430 Le Tampon

tél : 06 92 29 44 87

fax : 02 62 57 95 40

email : frederic.miranville@univ-reunion.fr

# List of figures

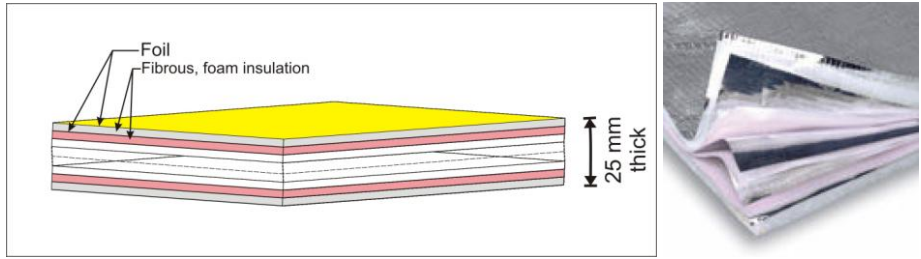

Figure 1: Multireflective insulation as the new technology for reflective insulation

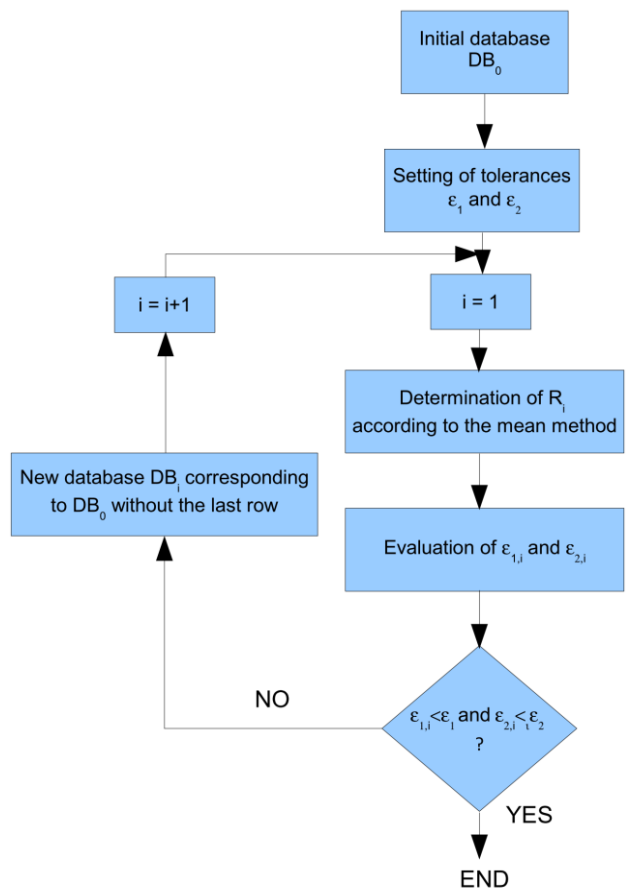

Figure 2: Synoptic of the process of calculation of the thermal resistance implemented in *ISOLAB*

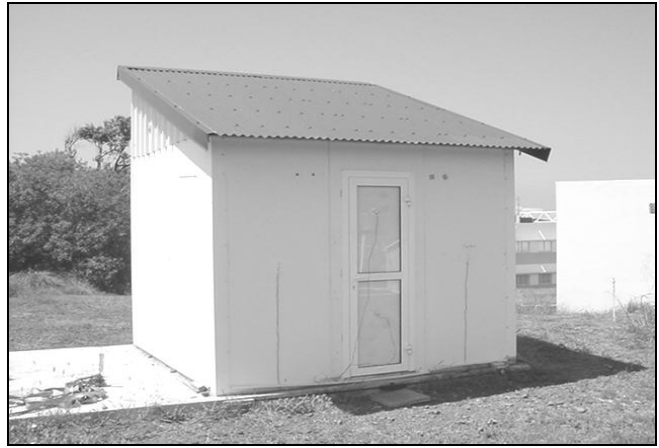

Figure 3: The LGI test cell

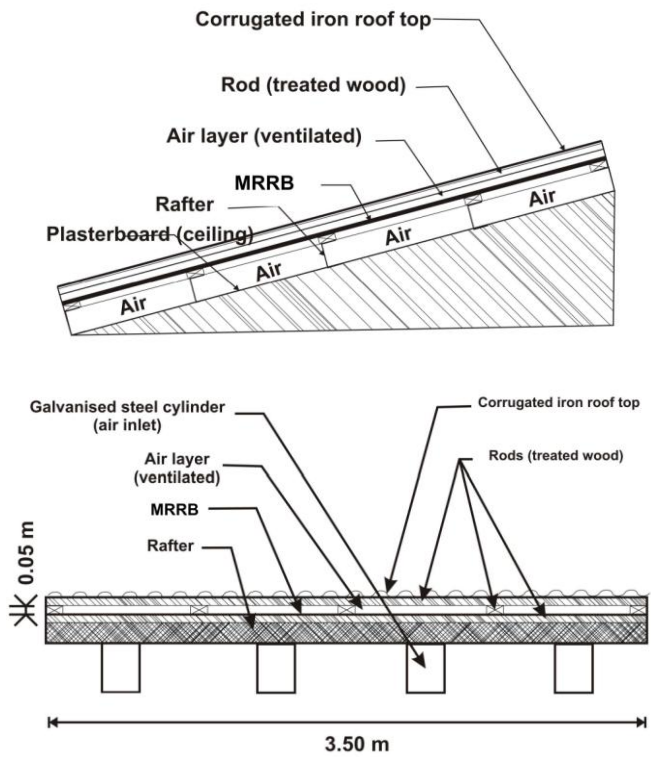

Figure 4: Section and front view of the specific roof installed for the study of MRRB (Multi-Reflective

Radiant Barriers)

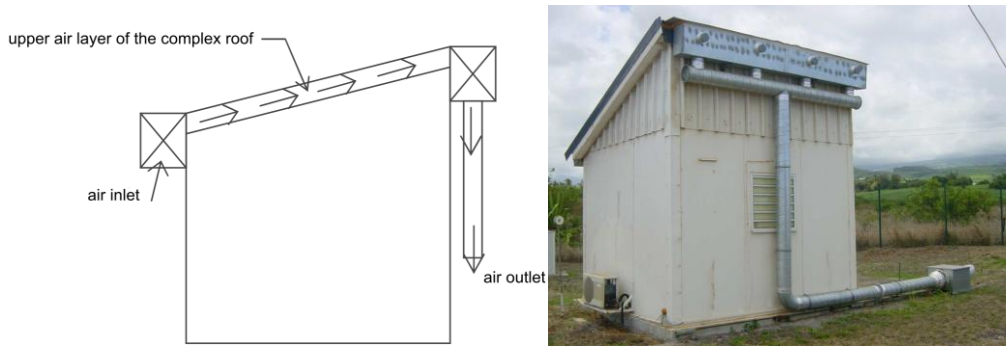

Figure 5: Section view and photo of the device for the mechanical ventilation of the upper air layer of

the roof of the LGI cell

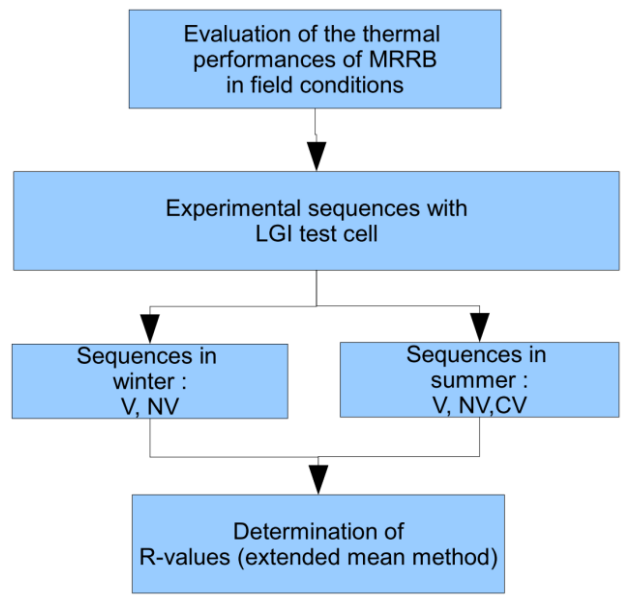

Figure 6: Overview of the exploitation method of measurements

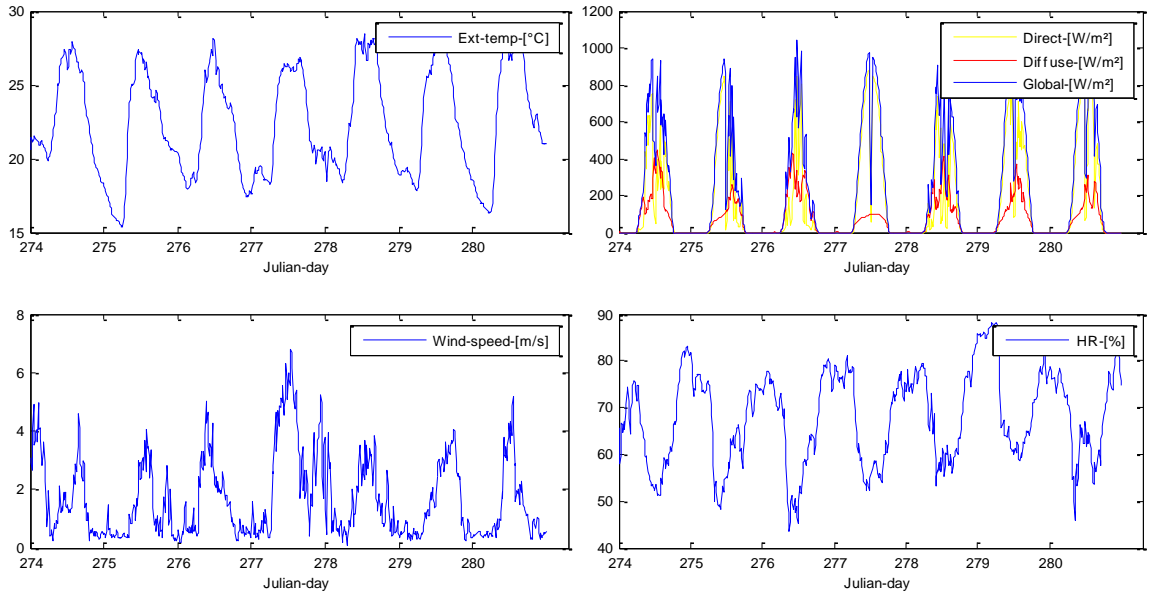

Figure 7: Climatic conditions for the winter experimental sequence (Direct, Diffuse and Global indicate

solar radiation)

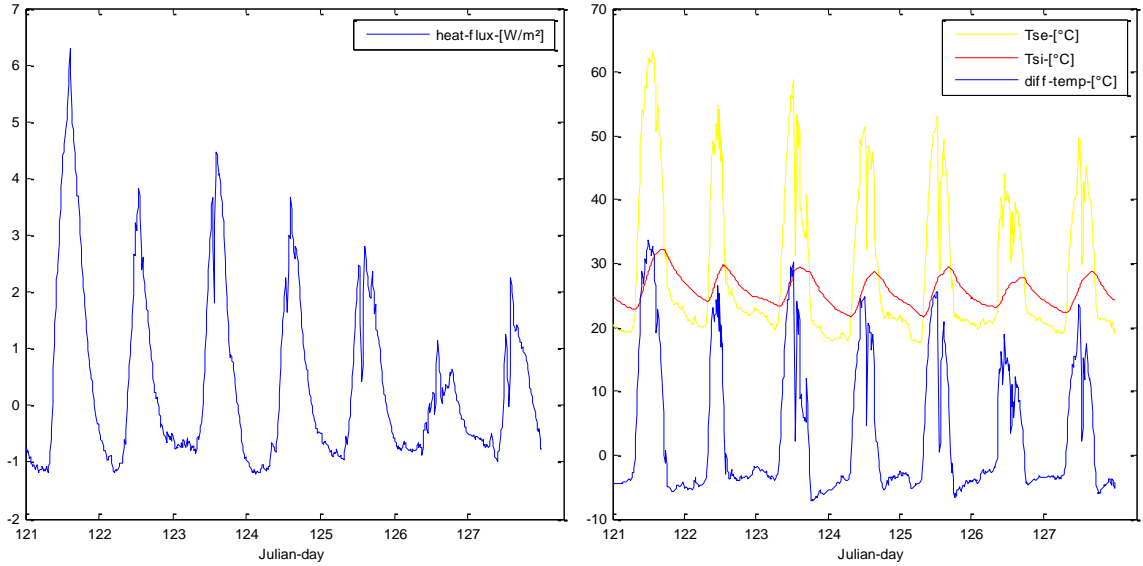

Figure 8: Heat flux through the roof and temperature difference between the two boundaries (Tse:

temperature of the exterior surface of the roof – Tsi: temperature of the interior surface of the roof) for

the winter period - Natural ventilation case

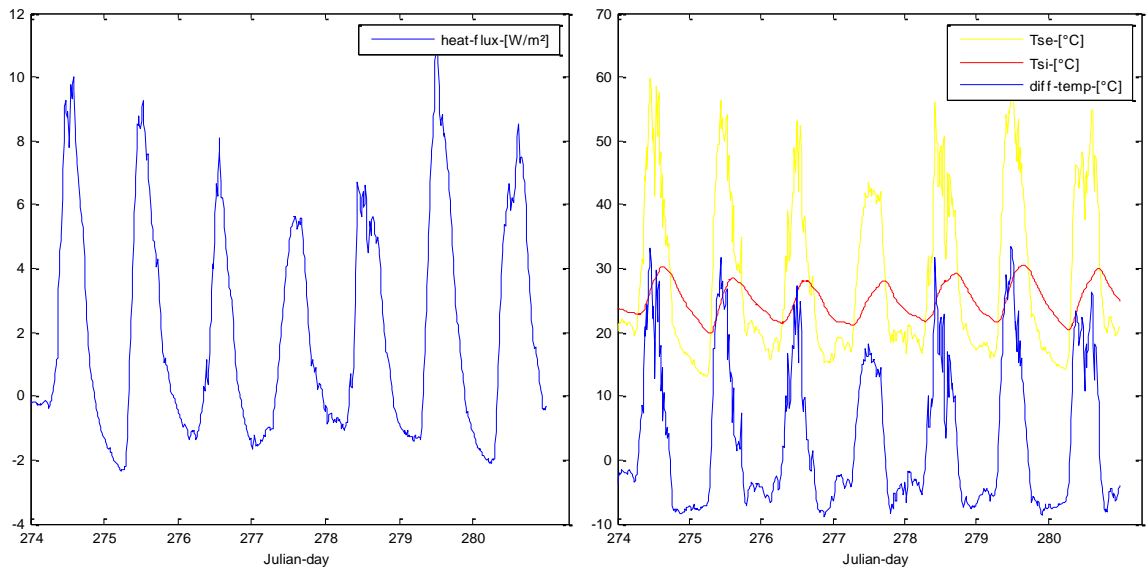

Figure 9: Heat flux through the roof and temperature difference between the two boundaries (Tse: temperature of the exterior face of the roof – Tsi: temperature of the interior face of the roof) of the roof for the winter period - No ventilation case

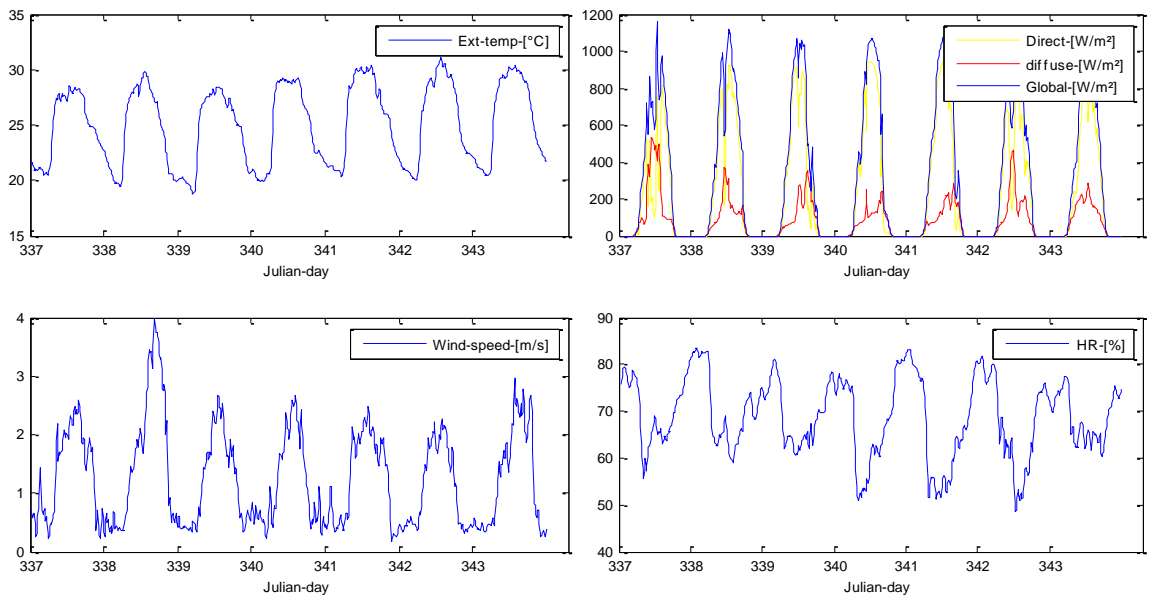

Figure 10: Climatic conditions for the summer experimental sequence (Direct, Diffuse and Global indicate solar radiation)

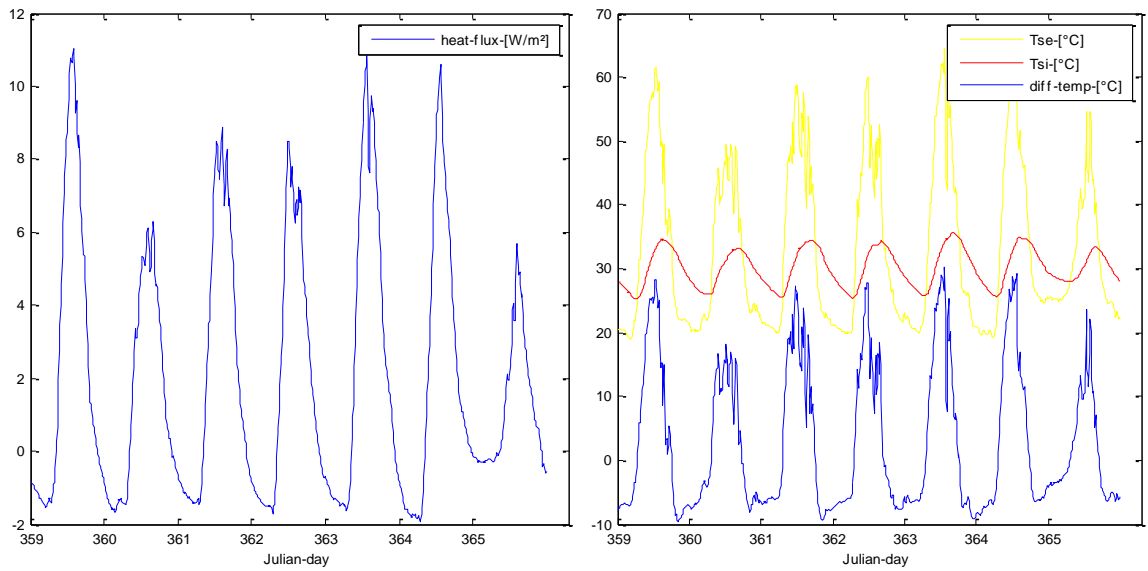

Figure 11: Heat flux through the roof and temperature difference between the two boundaries (Tse: temperature of the exterior face of the roof – Tsi: temperature of the interior face of the roof) for the summer period - Natural ventilation case

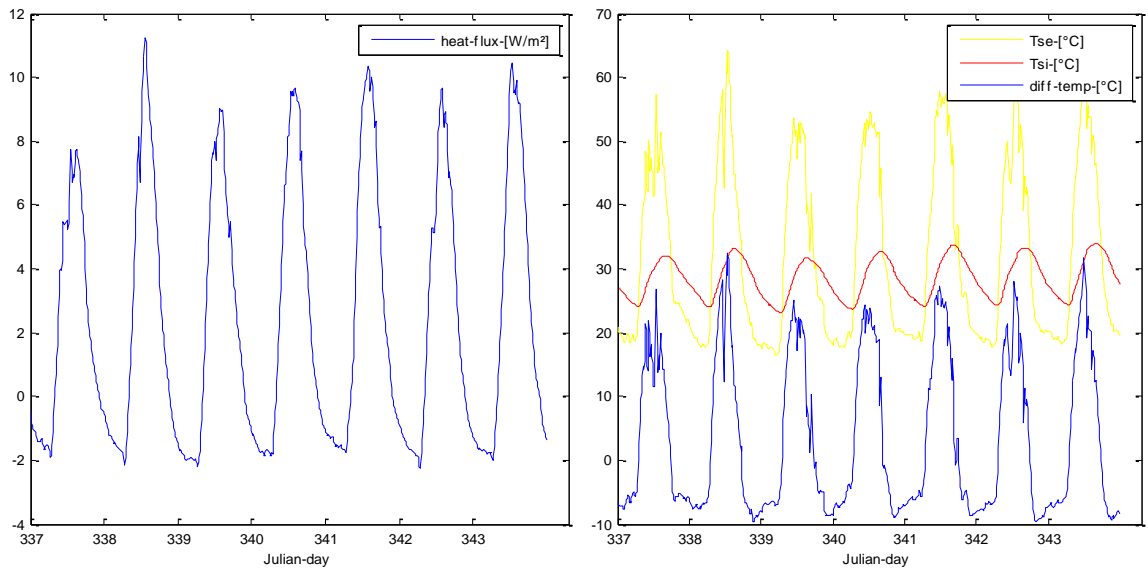

Figure 12: Heat flux through the roof and temperature difference between the two boundaries (Tse: temperature of the exterior face of the roof – Tsi: temperature of the interior face of the roof) for the summer period - No ventilation case

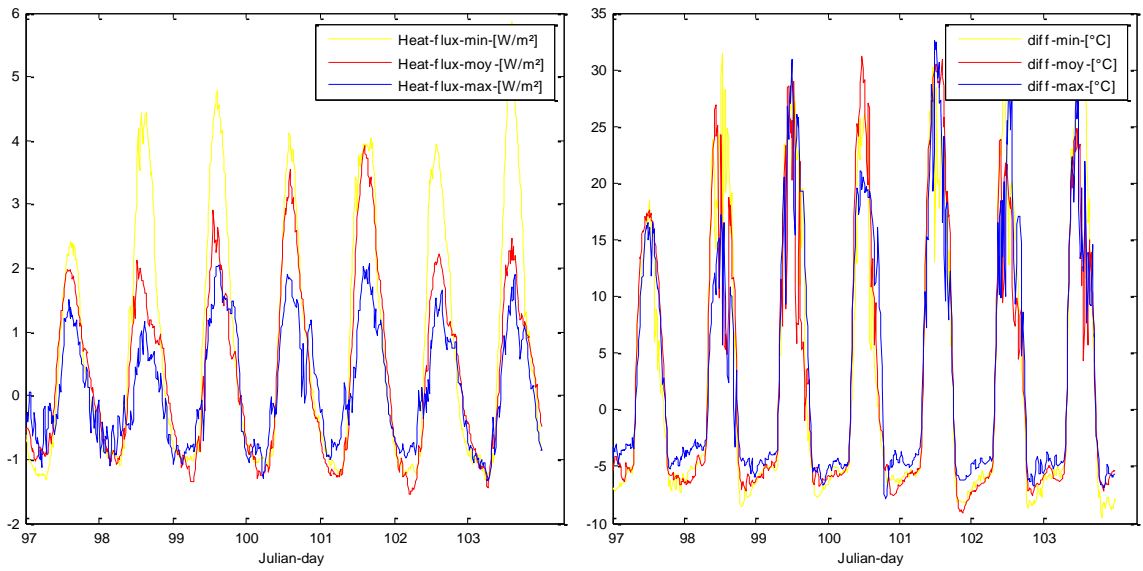

Figure 13: Heat flux and temperatures differences evolutions through the roof, according to the airflow rate in the upper air layer

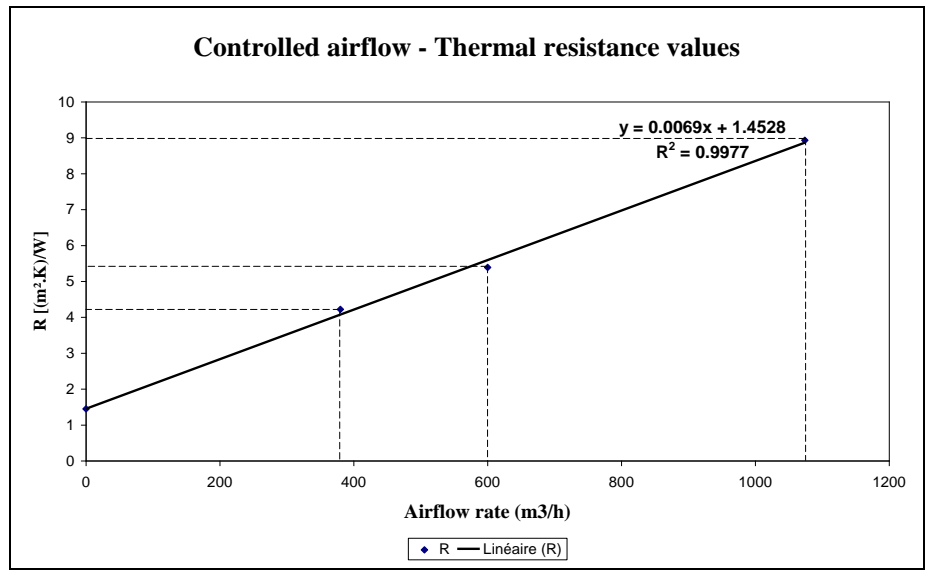

Figure 14: Evolution of thermal resistance according to the airflow rate in the upper air layer of the roof

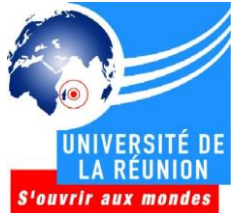
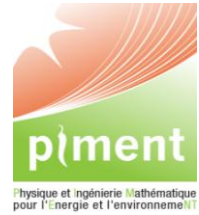

# Submission of manuscript to Energy and Buildings

## Evaluation of the thermal resistance of a roof-mounted multi-reflective radiant barrier for tropical and humid conditions:

### *Experimental study from field measurements*

Frédéric MIRANVILLE, Ali Hamada FAKRA, Stéphane GUICHARD, Harry BOYER,

Jean-Philippe PRAENE and Dimitri BIGOT

Contents:

- *List of tables*

Corresponding author:

**Pr. Frédéric MIRANVILLE**

Physics and Mathematical Engineering Laboratory for Energy and Environment (PIMENT)

Universiy of Reunion

117, rue du Général Ailleret

97430 Le Tampon

tél : 06 92 29 44 87

fax : 02 62 57 95 40

email : frederic.miranville@univ-reunion.fr

# List of tables

| Element | Composition | Remark(s) |
|---|---|---|
| Opaque vertical walls | Sandwich board 80mm thick cement-fibre / polyurethane / cement-fibre | |
| Window | Aluminium frame, 8 mm clear glass | Blind-type 0.8x0.8m |
| Glass door | Aluminium frame, 8mm clear glass | Glass in upper and lower parts, 0.7x2.2m |
| Roofing complex | Corrugated galvanised steel/air layer 100mm thick/RBS of 8mm thickness/air layer 16mm thick/plasterboard 8mm thick (inclination 20°) | RBS composed of aluminium faces and a polyethylene interface |
| Floor | Concrete slabs of thickness 80mm on 60 mm thick polystyrene | |

Table 1: Details of the construction of the LGI cell

| Experimental period | LGI test cell |
|---|---|
| Summer | Upper air layer ventilated (V) |
| | Upper air layer non ventilated (NV) |
| | Controlled ventilation of upper air layer (CV) |
| Winter | Upper air layer ventilated (V) |
| | Upper air layer non ventilated (NV) |

Table 2: Configurations of the test cells during the experimental period

| | Airflow speed | Airflow rate |
|---|---|---|
| Sequence 1 | 3.2 m/s | 380 m$^3$/h |
| Sequence 2 | 5.6 m/s | 600 m$^3$/h |
| Sequence 3 | 9.5 m/s | 1074 m$^3$/h |

Table 3: Results of the determination procedure of the airflow rate in the upper air layer of the LGI cell

| Case | Values of airflow rate | Pc |
|---|---|---|
| Airflow rate 1 | 380 m$^3$/h | **60%** |
| Airflow rate 2 | 600 m$^3$/h | **86%** |
| Airflow rate 3 | 1074 m$^3$/h | **104 %** |

Table 4: Percentage of reduction of heat flux through the roof, according to the airflow rate of the upper

air layer

| Airflow speed [m/s] | Airflow rate [m$^3$/h] | Resistance value [m².K/W] |
|---|---|---|
| 0 | 0 | 1.47 |
| 3.5 | 380 | 4.22 |
| 5.6 | 600 | 5.39 |
| 9.5 | 1074 | 8.93 |

Table 5: Thermal resistance values according to several airflow rates

| *R-values from LGI cell* | *Case with upper air layer naturally ventilated* | *Case without upper air layer ventilated* |
|---|---|---|
| Summer | 1.44 m$^2$.K.W$^{-1}$ | 1.47 m$^2$.K.W$^{-1}$ |
| Winter | 6.24 m$^2$.K.W$^{-1}$ | 1.66 m$^2$.K.W$^{-1}$ |

Table 6: R-values from field measurements using LGI cell